\documentclass[letterpaper]{article} 
\usepackage{aaai25}  
\usepackage{times}  
\usepackage{helvet}  
\usepackage{courier}  
\usepackage[hyphens]{url}  
\usepackage{graphicx} 
\usepackage{amsmath}
\usepackage{natbib}  
\usepackage{caption} 
\usepackage{booktabs} 
\usepackage{algorithm}
\usepackage{algorithmic}

\usepackage{newfloat}
\usepackage{listings}
\DeclareCaptionStyle{ruled}{labelfont=normalfont,labelsep=colon,strut=off} 
\floatstyle{ruled}
\newfloat{listing}{tb}{lst}{}
\floatname{listing}{Listing}
\newcommand{\domain}[1]{\textit{#1}}
\newcommand{\model}[1]{\textit{#1}}

\title{News Source Citing Patterns in AI Search Systems}
\author{
    Kai-Cheng Yang
}
\affiliations{
    Northeastern University, Boston, MA, USA\\
    yang3kc@gmail.com
}

\begin{document}

\maketitle

\begin{abstract}
AI-powered search systems are emerging as new information gatekeepers, fundamentally transforming how users access news and information.
Despite their growing influence, the citation patterns of these systems remain poorly understood.
We address this gap by analyzing data from the AI Search Arena, a head-to-head evaluation platform for AI search systems.
The dataset comprises over 24,000 conversations and 65,000 responses from models across three major providers: OpenAI, Perplexity, and Google.
Among the over 366,000 citations embedded in these responses, 9\% reference news sources.
We find that while models from different providers cite distinct news sources, they exhibit shared patterns in citation behavior.
News citations concentrate heavily among a small number of outlets and display a pronounced liberal bias, though low-credibility sources are rarely cited.
User preference analysis reveals that neither the political leaning nor the quality of cited news sources significantly influences user satisfaction.
These findings reveal significant challenges in current AI search systems and have important implications for their design and governance.
\end{abstract}

%
\begin{links}
    \link{Code}{https://github.com/yang3kc/ai_search_arena}
\end{links}

\section{Introduction}

AI-powered search systems are transforming how users access and consume information online.
Unlike traditional web search engines that return ranked lists of web pages, AI search systems synthesize information from multiple sources and present coherent, conversational responses augmented with citations to supporting evidence~\cite{xiong2024searchengineservicesmeet}.
Studies suggest that these systems lower the information access barrier~\cite{wu2020providing} and enable users to perform complex tasks more quickly~\cite{spatharioti2023comparingtraditionalllmbasedsearch,suri2024usegenerativesearchengines}.
Products such as ChatGPT with web search, Perplexity AI, and Google's AI search features have gained millions of users, demonstrating the potential of AI search systems to become mainstream.

Scholars have long conceptualized gatekeeping as the process of selecting and highlighting certain information while filtering out other content~\cite{barzilai2008toward,shoemaker2009gatekeeping}.
Originally applied to human editors and journalists, this concept has been extended to algorithmic systems that filter and prioritize information flows at scale~\cite{van2023algorithmic}.
AI search engines represent a new evolution of this phenomenon: rather than simply retrieving documents, they actively synthesize information and foreground specific sources.
As these systems increasingly serve as primary gateways to information, they exercise unprecedented control over which outlets reach users, effectively functioning as the new gatekeepers of the digital information ecosystem~\cite{kuai2025ai}.

However, this gatekeeping role raises critical questions about which sources these systems prioritize and how their citation patterns shape public discourse~\cite{memon2024searchenginespostchatgptgenerative}.
These concerns are not new since audits of traditional search engines and news aggregators have shown that cited news sources concentrate among a relatively small number of popular outlets and exhibit liberal bias~\cite{trielli2019search,fischer2020auditing}.
Built upon traditional information retrieval systems, AI search systems may amplify these issues through their complex and opaque mechanisms that can embed systematic biases in source selection~\cite{yang2025accuracy}.

Recent audits of popular AI search engines confirm these concerns.
Using simulated user queries, researchers have shown that tools like Microsoft Copilot tend to favor mainstream news outlets~\cite{brantner2025sourcing}.
These systems have also been found to surface more low-credibility sources than Google News and demonstrate left-leaning bias~\cite{li2024generativeaisearchengines,jaidka2025news}.
Given the contentious nature of the contemporary media landscape, AI systems' tendency to favor sources with particular characteristics could contribute to information bubbles and exacerbate information disparities.

While this emerging evidence provides valuable insights, critical gaps remain.
First, previous audits of AI search systems rely primarily on simulated user queries, which may not accurately reflect real-world usage patterns.
Second, these audits typically examine only one or a few AI search systems, limiting their representativeness and preventing the identification of cross-model patterns.
Third, these studies lack user preference data, making it difficult to determine which factors drive user satisfaction with search results.

To address these gaps, we conduct a large-scale empirical analysis of news citation patterns in AI search systems using the AI Search Arena dataset.
The dataset contains over 366,000 citations from conversations between users and 12 AI search models from three major providers: OpenAI, Perplexity, and Google.
The dataset is collected through a head-to-head evaluation setup where users are presented with responses from different AI search models for the same query and asked to choose the better response.
This dataset enables us to examine the citation patterns of AI search systems in real-world usage scenarios.
Specifically, we focus on two key research questions:

\begin{itemize}
    \item \textbf{RQ1}: Do different AI search systems exhibit systematic patterns in their news citation behavior, and if so, what factors are associated with these patterns?
    \item \textbf{RQ2}: How do the characteristics of cited news sources relate to user preferences for AI search responses?
\end{itemize}

Our analysis reveals the following key findings.
First, AI search models from the same provider exhibit similar citation patterns, while differences between providers are more pronounced.
Second, all models demonstrate significant citation concentration, with a small number of news outlets accounting for a disproportionate share of citations.
Third, we observe consistent left-leaning political bias across all AI search systems, despite their general preference for high-quality sources.
Notably, however, user preferences for AI search responses show no significant association with either the political orientation or quality of cited news sources.

Our findings confirm the gatekeeping role of AI search systems and reveal potential issues in their news-citing behaviors.
These findings have important implications for the design of AI search systems and their role in shaping public access to information.

\section{Related Work}

\subsection{Sources Cited by Traditional Search Engines}

AI search systems are typically implemented as a combination of large language models (LLMs) and web search engines~\cite{xiong2024searchengineservicesmeet}.
Therefore, audits of traditional search engines and news aggregators are particularly relevant to our study.
Previous research reveals a consistent pattern: cited news concentrates heavily on a relatively small number of popular outlets in traditional search engines~\cite{trielli2019search,fischer2020auditing}.
This concentration persists regardless of user identities and personalization settings~\cite{trielli2022partisan}.
While low-credibility sources do appear in search results, they remain relatively rare~\cite{robertson2023users}.

Regarding political bias, research from organizations like AllSides\footnote{https://www.allsides.com/blog/google-news-shows-strong-political-bias-allsides-analysis} and academic studies~\cite{robertson2018auditing,robertson2023users} suggest that Google tends to display more left-leaning sources in search results and top stories.
This tendency appears consistent across users with different political leanings and personalization settings.
However, user queries themselves vary in their political orientation, leading to results with different degrees of political tendencies~\cite{robertson2018auditing,tong2025unite}.
In this study, we examine whether similar patterns emerge in AI search systems.

\subsection{Sources Cited by AI Search Systems}

The rise of AI search systems has drawn significant attention from the academic community, with many auditing studies emerging in recent years.
A common approach involves creating queries covering various topics and then examining responses from different AI search tools.

For instance, \citet{brantner2025sourcing} focus on Microsoft Copilot during the run-up to Taiwan's 2024 presidential election to examine what content it surfaces and how accurately it cites sources.
The authors also test the same queries in different languages to assess how the model responds across linguistic contexts.
Their results suggest that Copilot relies heavily on professional news outlets and skews heavily toward UK- and US-based English sources across all languages.
While deliberate disinformation is rare, they find that many professional news citations are mis-summarized or misattributed, and Bing links are often broken, making misinformation a byproduct of summary and technical errors.

Similarly, \citet{li2024generativeaisearchengines} audit three AI search systems---ChatGPT, Bing Chat, and Perplexity---by collecting over 1,000 answers to 48 public-interest queries.
The authors analyze the cited sources and show that they are dominated by US commercial and news sites with scant academic or governmental material.
They warn that these patterns may amplify confirmation bias and commercial/geographic skew, recommending stricter source-quality filters and greater transparency in generative AI search.

\citet{jaidka2025news} adopt a comparative approach by sending queries to an AI search tool and Google News.
The authors use over 80,000 queries across 11 election-year countries covering various topics.
Unlike other studies, the AI search tool being tested combines OpenAI's GPT-4o model with a news API, giving the authors better control over the underlying mechanisms.
They show that the AI search tool surfaces significantly less relevant and more low-quality content while being only slightly less left-leaning overall than Google News, with the quality gap widest for right-party queries and deeper result pages.

In their qualitative study, \citet{venkit2025searchengines} recruit 21 expert users to compare AI search systems to traditional search engines.
The experts commonly identify problems in the AI search systems, such as misattribution and misrepresentation of cited sources, missing citations for certain claims, and a lack of transparency in source selection.
In particular, these experts express general distrust toward certain sources cited by AI search systems, especially forums, blogs, and opinion pieces.

These studies provide valuable insights into the citation patterns of AI search systems.
In this study, we contribute to this line of research by examining a large-scale dataset from real-world interactions between users and AI search systems and covering a wide range of AI search models from different providers.

\section{Data and Methods}

\subsection{AI Search Arena Dataset}

Our analysis is based on the AI Search Arena dataset~\citep{miroyan2025searcharena}, which contains 24,069 conversations between users and AI search systems.\footnote{The dataset can be downloaded at https://huggingface.co/datasets/lmarena-ai/search-arena-24k}
These conversations were collected from March to May 2025, spanning seven weeks.
The dataset captures real-world usage patterns through a platform where users can interact with different AI search models and compare their responses in head-to-head evaluations.
Each conversation includes the user's query and responses from two AI models.
Some conversations include multiple turns of interaction.
Approximately half of the conversations are rated by users, who select a winner or declare a tie.
Some conversations also include information about the language of the conversation and the country of the user.

We extracted 366,087 citations (URLs) from responses generated by 12 AI search models from three providers: OpenAI, Perplexity, and Google.
The models analyzed include variations with different capabilities and settings.
For each citation, we extract the domain (e.g., \domain{nytimes.com} for The New York Times) for further analysis.
Summary statistics of the dataset are provided in the Appendix.

\subsection{News Outlet Identification}

The 366,087 citations link to 83,533 unique domains.
To understand the news citation patterns, we identify the news outlets leveraging a comprehensive list of news domains~\citep{yang2025newsdomains}.
This list is compiled from multiple sources, including NewsGuard,\footnote{https://www.newsguardtech.com} MBFC (Media Bias/Fact Check),\footnote{https://mediabiasfactcheck.com} ABYZ,\footnote{http://www.abyznewslinks.com} and Media Cloud.\footnote{https://mediacloud.org}
The list also incorporates news domains from datasets curated by academics, including \citet{robertson2018auditing}, \citet{cronin2023null}, \citet{le2022understanding}, \citet{fischer2020auditing}, and \citet{horne2022nela}.
We identify 1,795 (2.1\%) unique domains as news outlets, which account for 32,865 (9.0\%) of all citations.
In the Appendix, we also provide the percentage of citations for other domain types, such as social media and technology, to provide additional context.

\subsection{News Outlet Political Leaning}

We leverage the DomainDemo dataset to obtain political leaning scores for news outlets~\cite{yang2025domaindemo}.
This dataset is derived from a panel of over 1.5 million Twitter users whose accounts are matched to their voter registration records, linking users' source-sharing behaviors to their political affiliations.
Based on this information, the authors generate an audience-based metric that quantifies the political leaning of news outlets for over 129,000 domains.

The resulting political leaning scores range from -1 to +1, with -1 indicating sources predominantly shared by Democratic users and +1 indicating sources predominantly shared by Republican users.
The authors validated this metric by comparing it with existing datasets from academic studies, such as \citet{bakshy2015exposure}, \citet{eady2020news}, \citet{buntain2023measuring},
and organizations, such as MBFC and AllSides, finding high levels of agreement.

Merging the ratings from the DomainDemo dataset with our analytical sample, we obtain political leaning scores for 1,646 (91.7\%) unique domains, which account for 98.6\% of all news citations.
We use thresholds of -0.33 and 0.33 to classify news domains as ``left-leaning,'' ``center,'' and ``right-leaning.''
Domains without political leaning scores are labeled as ``unknown political leaning.''

\subsection{News Outlet Quality}

We adopt the news source quality ratings generated by \citet{lin2023high}.
These ratings are compiled from multiple sources, including NewsGuard, MBFC, Iffy index of unreliable sources,\footnote{https://iffy.news} and two lists generated by professional fact-checkers and researchers~\cite{pennycook2019fighting,lasser2022social}.
The authors aggregate these ratings to produce a comprehensive list covering 11,520 websites.
The final ratings are obtained through principal component (PCA) analysis, and the scores are normalized to range from 0 to 1, with higher scores indicating higher quality.

Merging the ratings from \citet{lin2023high} with our analytical sample leads to quality ratings for 1,283 (71.5\%) unique domains, which account for 80.8\% of all news citations.
We use a threshold of 0.5 to classify news domains as either ``high-quality'' or ``low-quality.''
News domains without quality rating scores are labeled as ``unknown quality.''

\subsection{Citation Concentration Measure}

To measure the concentration of citations across unique domains, we use the Gini index $G$, which is calculated as:
\begin{equation}
    G = \frac{1}{2 U^2 \bar{x}} \sum_{i=1}^U \sum_{j=1}^U |x_i - x_j| \text{,}
\end{equation}
where $U$ is the number of unique domains, $x_i$ is the number of citations to the $i$-th domain, and $\bar{x} = \sum_{i=1}^U x_i / U$ is the average number of citations per domain.
The Gini index ranges from 0 to 1, where 0 indicates perfect equality, and 1 indicates perfect inequality.

\subsection{User Question Characteristics}

To understand the association between news citation patterns and user questions, we extract rich linguistic features from the user questions.
We calculate sentence embeddings using SentenceTransformers with the ``all-MiniLM-L6-v2'' model~\citep{reimers2019sentencebertsentenceembeddingsusing}, which capture the semantic characteristics of the questions.

Since embedding vectors are cryptic, we also include more interpretable features.
First, we classify the questions by their intent.
The intent taxonomy is provided by \citet{miroyan2025searcharena} and includes the following categories: ``factual lookup,'' ``information synthesis,'' ``analysis,'' ``recommendation,'' ``explanation,'' ``creative generation,'' ``guidance,'' ``text processing,'' and ``other.''

Second, we perform topic modeling on the questions using BERTopic~\citep{grootendorst2022bertopicneuraltopicmodeling}.
We follow recommended best practices by using the ``all-MiniLM-L6-v2'' model to generate sentence embeddings, UMAP to reduce dimensions, and HDBSCAN to cluster topics.
This procedure yields 10 topics, and we manually generate labels such as ``AI models and technology,'' ``news updates,'' and ``sports and entertainment'' for them.
Since individual questions might be associated with multiple topics, we calculate the topic distribution for each question as its topic features.
This distribution captures the similarity of the question to each topic.
The full list of representative keywords of each topic and their labels are provided in the Appendix.

\subsection{Regression Analysis}

To understand the factors associated with news citation patterns, we conduct regression analyses using each user question-AI response pair as the unit of analysis.
We focus on a subset of the dataset that meets two criteria.
First, we include only questions labeled as English to ensure consistent linguistic feature extraction.
Second, we restrict our analysis to responses containing at least one news citation, as our primary interest lies in understanding news citation patterns.
These criteria yield 9,098 observations.

The dependent variables for our linear regression models are the percentages of left-leaning, right-leaning, high-quality, and low-quality news citations among all news citations in each response (rather than among all citations).
This approach allows us to examine the composition of news sources when AI models choose to cite news outlets.

Our regression models include the following independent variables.
At the conversation level, we include the total number of conversation turns and the country/region of the client.
We create dummy variables for the 10 most frequent countries/regions in the dataset and code all other countries/regions as ``other.''
Conversations with missing country/region information are coded as ``unknown.''
The US serves as the reference category.

At the response level, we include model family dummy variables, with Perplexity as the reference category.
We also include the turn number for each response, the number of citations, and the percentage of news citations among all citations.
Additionally, we include the word count of the response, which is log-transformed and converted to Z-scores to address distributional skewness.

At the question level, we include the word count of each question, which is log-transformed and converted to Z-scores.
We apply PCA to reduce the aforementioned sentence embedding with 384 dimensions to 10 principal components and include them in the regression models as control variables.
Intent features are encoded as dummy variables with ``recommendation'' serving as the reference category.
Topic distribution features are standardized to Z-scores and incorporated into the regression models.

\subsection{User Preference Analysis}

To understand the relationship between news citation patterns and user preferences, we focus on conversations where both responses cite at least one news source.
We also exclude conversations with no user judgment or with a tie, leading to 1,534 head-to-head comparisons.

We employ the Bradley-Terry model to estimate the association between different features and user choices~\cite{bradley1952rank,chiang2024chatbotarena}.
The model assumes that the probability of preferring response A over B depends on the difference in their latent quality scores, which are modeled as linear combinations of observed features.
We estimate coefficients using maximum likelihood estimation and assess 95\% confidence intervals and statistical significance through bootstrap resampling with 1,000 replications.

We calculate the proportion of news citations among all citations as the control variable.
We then calculate the proportions of left-leaning, center, and right-leaning news citations, and the proportions of high-quality and low-quality news citations among news citations to test their association with user preferences.
To control for additional confounding factors, we include response word count and total citation count, both of which have been identified as key determinants of user preferences~\cite{miroyan2025searcharena}.
Since user judgments are provided at the conversation level, we average all response-level variables across turns when a conversation involves multiple rounds of exchanges to produce final conversation-level variables.
The difference in style features between the two AI models is transformed into Z-scores before being fed into the Bradley-Terry model.

\section{Results}

\subsection{Basic News Citations Characteristics}

\begin{figure}
    \centering
    \includegraphics[width=0.85\columnwidth]{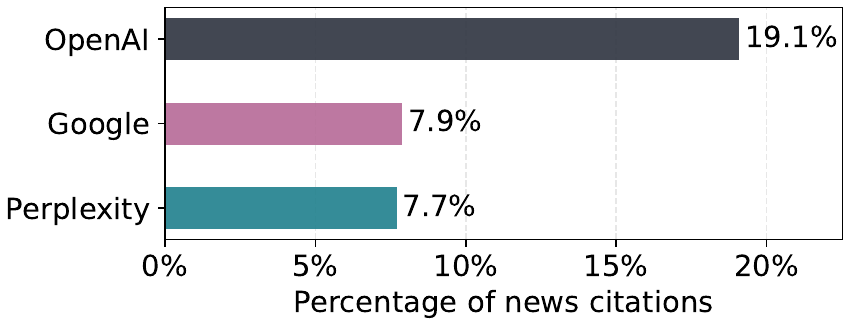}
    \caption{News citation rates among all citations across model families.}
    \label{fig:model_family_news_citation_rates}
\end{figure}

Our analysis reveals that models within the same family exhibit similar citation patterns.
Specifically, models from the same family tend to cite similar proportions of news sources among all citations.
A detailed breakdown by individual model is provided in the Appendix, whereas Figure~\ref{fig:model_family_news_citation_rates} displays the news citation rates across model families.
We find that OpenAI models consistently cite news sources at higher rates than their counterparts, while Google and Perplexity models show comparable citation frequencies.

The similarity extends beyond citation rates to the specific news domains cited.
Our cosine similarity analysis of news citations (detailed in the Appendix) shows that models within the same family achieve similarity scores ranging from 0.82 to 0.99.
In contrast, cross-family similarities are substantially lower (ranging from 0.11 to 0.58).
These findings support our decision to aggregate subsequent analyses at the model family level, focusing on inter-family differences rather than intra-family variations.

\subsection{Frequent News Sources and Concentration}

\begin{table*}[htbp]
\centering
\caption{
    Top 20 most frequent news sources by the model family with political leaning and quality.
    Political leaning (columns ``L''): L=left-leaning, C=center, R=right-leaning, U=unknown.
    Quality (columns ``Q''): H=high-quality, L=low-quality, U=unknown.
}
\label{tab:top_news_sources}
\begin{tabular}{lrcc|lrcc|lrcc}
\toprule
\multicolumn{4}{l|}{\textbf{OpenAI}} & \multicolumn{4}{l|}{\textbf{Google}} & \multicolumn{4}{l}{\textbf{Perplexity}} \\
\midrule
Domain & \% & L & Q & Domain & \% & L & Q & Domain & \% & L & Q \\
\midrule
reuters.com & 22.8 & C & H & indiatimes.com & 4.3 & C & L & bbc.com & 3.2 & C & H \\
apnews.com & 12.2 & C & H & forbes.com & 2.5 & C & H & yahoo.com & 2.7 & C & H \\
ft.com & 7.0 & C & H & aljazeera.com & 2.5 & L & H & 163.com & 2.0 & C & U \\
axios.com & 6.7 & C & H & cbsnews.com & 2.0 & C & H & sohu.com & 2.0 & C & L \\
time.com & 3.9 & C & H & apnews.com & 2.0 & C & H & nytimes.com & 1.7 & C & H \\
elpais.com & 2.5 & L & H & techradar.com & 1.9 & C & H & cnn.com & 1.6 & C & H \\
as.com & 2.4 & C & U & healthline.com & 1.7 & C & H & reuters.com & 1.5 & C & H \\
theatlantic.com & 1.5 & L & H & 163.com & 1.5 & C & U & sina.com.cn & 1.5 & L & U \\
techradar.com & 1.1 & C & H & cbr.com & 1.4 & C & U & espn.com & 1.4 & C & H \\
lemonde.fr & 0.8 & L & H & yahoo.com & 1.3 & C & H & forbes.com & 1.3 & C & H \\
sohu.com & 0.8 & C & L & screenrant.com & 1.3 & C & U & rbc.ru & 1.3 & L & U \\
bbc.com & 0.8 & C & H & zdnet.com & 1.2 & C & H & cbsnews.com & 1.2 & C & H \\
knowyourmeme.com & 0.7 & L & H & sina.com.cn & 1.1 & L & U & indiatimes.com & 1.0 & C & L \\
indiatimes.com & 0.7 & C & L & pcmag.com & 1.1 & C & H & apnews.com & 0.9 & C & H \\
theguardian.com & 0.6 & C & H & hindustantimes.com & 1.1 & C & H & techradar.com & 0.9 & C & H \\
forbes.com & 0.6 & C & H & medicalnewstoday.com & 1.0 & C & H & screenrant.com & 0.9 & C & U \\
rbc.ru & 0.6 & L & U & livemint.com & 1.0 & C & U & techcrunch.com & 0.9 & C & H \\
healthline.com & 0.5 & C & H & pbs.org & 1.0 & L & H & businessinsider.com & 0.9 & C & H \\
usnews.com & 0.5 & C & H & thehindu.com & 1.0 & L & H & pbs.org & 0.8 & L & H \\
businessinsider.com & 0.4 & C & H & cnet.com & 0.9 & C & H & cnbc.com & 0.8 & C & H \\
\midrule
Total & 67.3 &  &  & Total & 31.9 &  &  & Total & 28.5 &  &  \\
\bottomrule
\end{tabular}
\end{table*}

Table~\ref{tab:top_news_sources} presents the top 20 most frequently cited news sources by model family, illustrating the distinct citation preferences across different AI systems.
Each model family exhibits clear preferences for specific domains.
For instance, OpenAI models most frequently cite \domain{reuters.com}, Google models favor \domain{indiatimes.com}, and Perplexity models prefer \domain{bbc.com}.
While some domains like \domain{apnews.com} appear across multiple families, the overall patterns align with our similarity analysis showing strong inter-family differences.
To further characterize these differences, we apply the log-odds ratio informative Dirichlet prior method~\citep{monroe2008fightin} in the Appendix to identify overrepresented news sources for each model family.
This approach down-weights commonly cited domains and isolates sources that are uniquely preferred by each model family.

Table~\ref{tab:top_news_sources} also reveals a highly concentrated citation pattern.
For example, OpenAI models cite \domain{reuters.com} and \domain{apnews.com} at rates of 22.8\% and 12.2\% of all citations, respectively.
Consequently, the top 20 most frequent news sources account for 67.3\% of all citations for OpenAI models.
While Google and Perplexity models show less severe concentration, their top 20 sources still represent 31.9\% and 28.5\% of all citations, respectively.

\begin{figure}
    \centering
    \includegraphics[width=0.8\columnwidth]{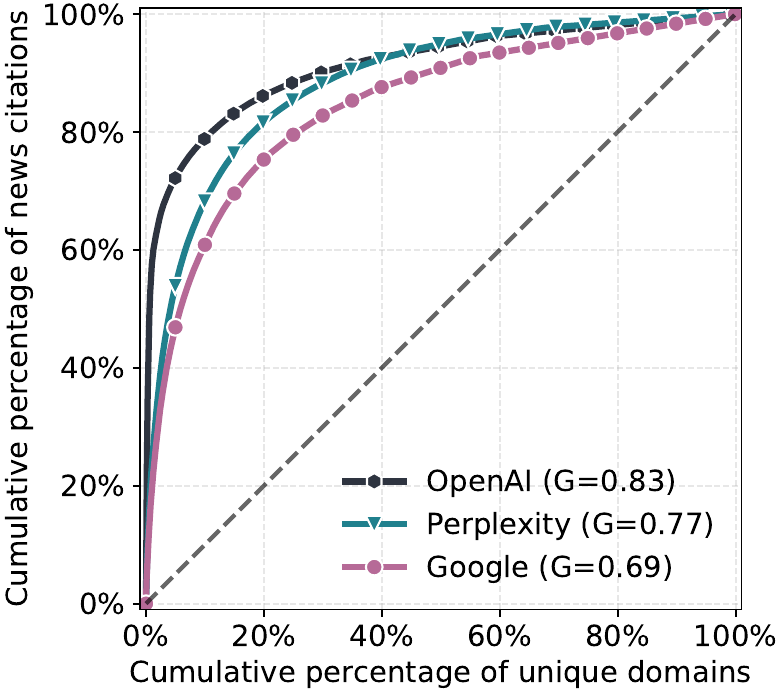}
    \caption{
    News citation Lorenz curves across model families.
    The diagonal line represents perfect equality.
    The Gini coefficients are annotated in the legend.
    }
    \label{fig:news_citation_lorenz_curves_families}
\end{figure}

To characterize the concentration of news citations across all news domains, we plot the Lorenz curves of the news citation rates across model families in Figure~\ref{fig:news_citation_lorenz_curves_families}.
We also calculate the Gini coefficients for each model family.
We can see that OpenAI models have the highest level of inequality (G=0.83) in their citation patterns, followed by Perplexity models (G=0.77) and Google models (G=0.69).

Comparing the results in Table~\ref{tab:top_news_sources} with the Gini coefficients reveals an apparent inconsistency.
While Google models show higher concentration among their top 20 news sources than Perplexity models (31.9\% vs. 28.5\%), Perplexity models exhibit higher overall inequality (G=0.77) than Google models (G=0.69).
This pattern suggests that Google models' concentration is driven by heavy reliance on their most frequently cited sources, whereas Perplexity models display more dispersed citation patterns across a broader range of domains.
A key contributing factor is that Perplexity models cite substantially more unique news sources (1,430) compared to Google (881) and OpenAI (707) models.
This long tail of domains cited only once or twice contributes to the higher Gini coefficients observed for Perplexity models, despite their lower concentration among top sources.

\subsection{Political Leaning and Quality of News Sources}

\begin{figure}[t]
    \centering
    \includegraphics[width=0.9\columnwidth]{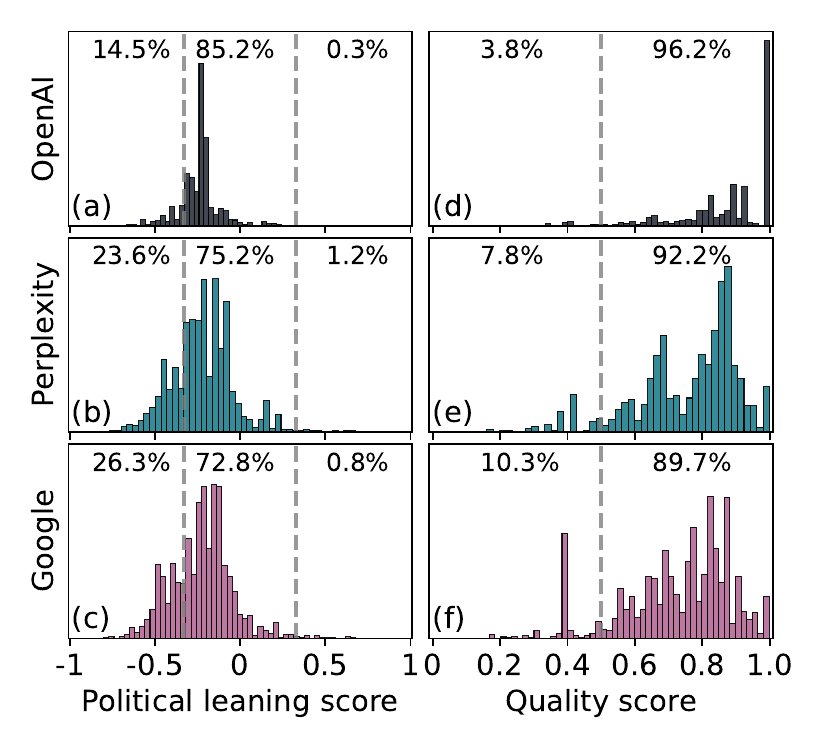}
    \caption{
    Distribution of political leaning and quality scores of news sources across model families.
    The dashed lines indicate the thresholds for political leaning and quality scores described in the main text.
    The percentages of sources in different categories are annotated in the figure.
    }
    \label{fig:model_family_leaning_quality_distribution}
\end{figure}

To assess the political leaning of news sources cited by different model families, we examine the distribution of source political leaning scores in Figure~\ref{fig:model_family_leaning_quality_distribution}.
Kolmogorov-Smirnov tests reveal that all pairwise distributions differ significantly ($p < 0.001$) even after correcting for multiple comparisons.
Applying the -0.33 and 0.33 thresholds, we classify the news domains in the citations into three categories: left-leaning, center, and right-leaning.
We find that OpenAI models demonstrate the highest proportion of center sources (85.2\%), compared to Google (75.2\%) and Perplexity (72.8\%) models.
For left-leaning sources, Google models show the highest citation rate at 26.3\%, followed by Perplexity models at 23.6\% and OpenAI models at 14.5\%.
Right-leaning sources represent a minimal fraction across all model families: 1.2\% for Perplexity, 0.8\% for Google, and 0.3\% for OpenAI models.
Despite these variations, all three model families exhibit a pronounced left-leaning bias in their news source citations.

Similarly, we show the distribution of source quality scores in Figure~\ref{fig:model_family_leaning_quality_distribution} as well.
Kolmogorov-Smirnov tests reveal that all pairwise distributions differ significantly ($p < 0.001$) even after correcting for multiple comparisons.
Applying a threshold of 0.5, OpenAI models demonstrate the highest proportion of high-quality sources (96.2\%), followed by Google models (92.2\%) and Perplexity models (89.7\%).

Table~\ref{tab:top_news_sources} also displays the quality and political leaning categories for the top 20 most frequently cited news sources by model family.
These domains are predominantly center and left-leaning sources and are overwhelmingly high-quality, consistent with the findings in Figure~\ref{fig:model_family_leaning_quality_distribution}.
The same pattern is observed for the overrepresented news sources as well (see the Appendix for details).

\subsection{Regression Analysis}

\begin{figure*}[t]
    \centering
    \includegraphics[width=\textwidth]{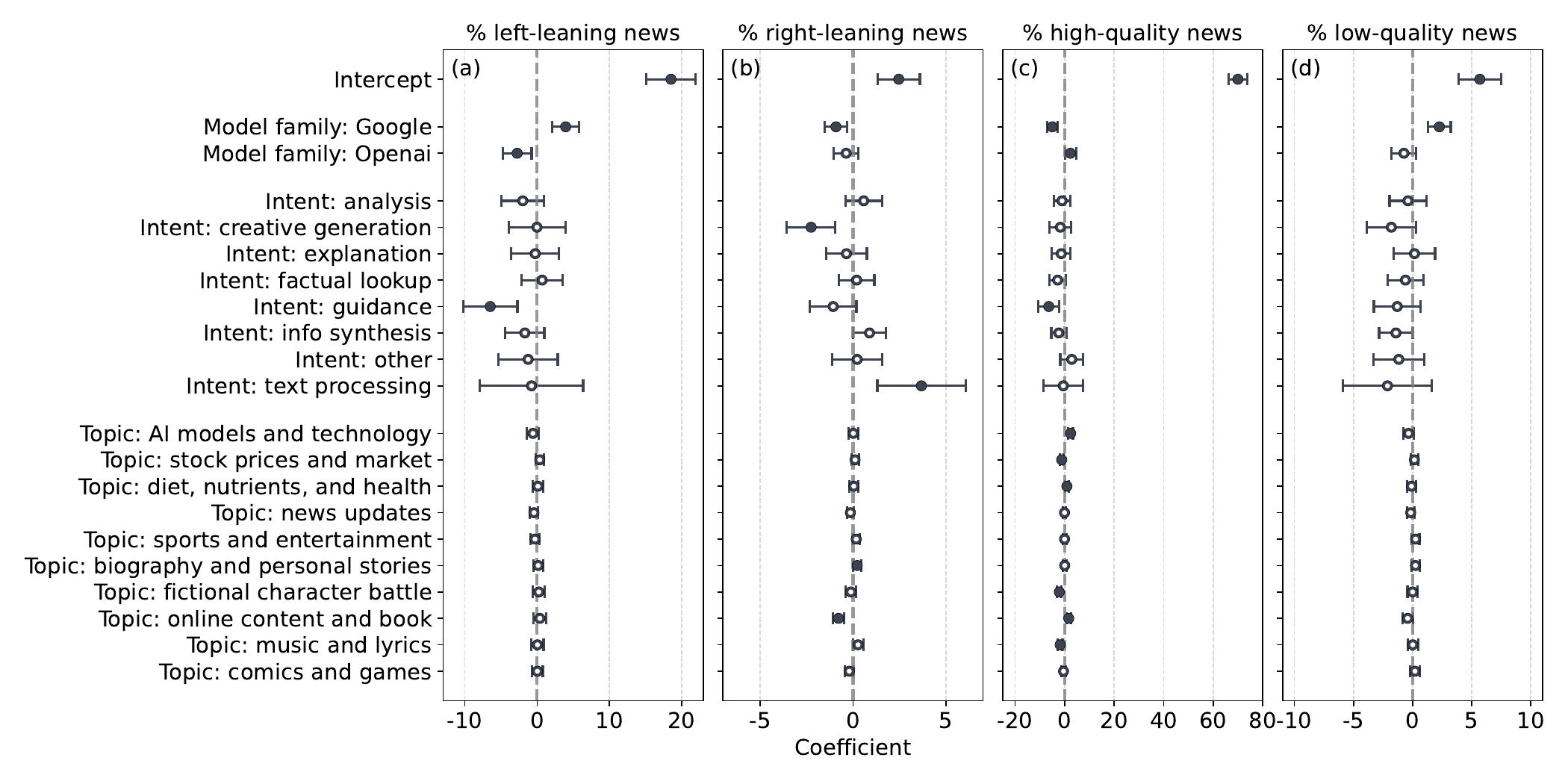}
    \caption{
    Coefficients for news source citation patterns in linear regression models.
    Each subfigure corresponds to a different dependent variable.
    Only coefficients for selected variables (intercepts, model family, question intent, and question topic) are displayed.
    Solid dots represent coefficients that are statistically significant at the 0.05 level, while hollow dots represent non-significant coefficients.
    Error bars indicate 95\% confidence intervals.
    Coefficients represent the percentage point change in the dependent variable for a one-unit increase in the independent variable.
    The model family and intent variables are binary indicators, with Perplexity and ``recommendation'' serving as reference categories.
    The topic variables represent the topic distribution of each question and are standardized to Z-scores.
    }
    \label{fig:regression_coefficients}
\end{figure*}

The patterns observed in Figure~\ref{fig:model_family_leaning_quality_distribution} may be confounded by various factors, such as the nature of users' questions.
To isolate the effects of different factors, we conduct linear regression analyses.
As described in the Data and Methods section, we regress the percentage of left-leaning, right-leaning, high-quality, and low-quality news citations among all news citations in each response on various conversation-level, question-level, and response-level variables.
Figure~\ref{fig:regression_coefficients} presents the regression coefficients for selected variables, with complete results available in the Appendix.
Since the dependent variables are percentages, the coefficients represent the percentage point change in the dependent variable for a one-unit increase in the independent variable.

The intercepts in the figures represent the baseline percentages of different types of news citations when all other variables are set to zero.
The intercept for left-leaning news citations is 18.5\%, substantially higher than the 2.5\% for right-leaning news, confirming the persistent left-leaning tendency in news citations.
Similarly, the intercept for high-quality news citations (69.9\%) far exceeds that for low-quality news (5.7\%), demonstrating a clear preference for high-quality news sources.

Examining model family differences, we find that compared to Perplexity models, Google models cite significantly more left-leaning and low-quality news sources while citing fewer right-leaning and high-quality sources.
Conversely, OpenAI models cite fewer left-leaning sources but more high-quality sources relative to Perplexity.
These regression results confirm the patterns observed in Figure~\ref{fig:model_family_leaning_quality_distribution}, demonstrating that the inter-family differences persist even after controlling for potential confounding factors.
Moreover, the coefficients for the model family variables are relatively small compared to the intercepts, suggesting that the tendency to cite left-leaning and high-quality sources is consistent across all model families.

Regarding question intent, questions in the ``guidance'' category are associated with decreased citations to both left-leaning and high-quality news sources.
Additionally, ``creative generation'' questions show decreased citations to right-leaning news sources, while ``text processing'' questions exhibit increased citations to right-leaning sources.

The patterns for the question topics are more complex.
It is worth noting that these variables represent the topic distribution converted to Z-scores, so the unit of change is one standard deviation.
We see no statistically significant associations of these variables with left-leaning and low-quality news citations.
However, for right-leaning news citations and high-quality news citations, there are some significant associations.
For instance, the topic ``online content and book'' is associated with decreased citations to right-leaning news but increased citations to high-quality news.

For clarity, we omit coefficients for several variables from the figures, as they fall outside our primary focus or are difficult to interpret.
Nevertheless, some of these omitted variables remain statistically significant.
For example, certain client country/region variables show significant associations with specific news citation patterns.
Complete regression results are available in the Appendix.

\subsection{User Preference Analysis}

\begin{figure}[t]
    \centering
    \includegraphics[width=0.9\columnwidth]{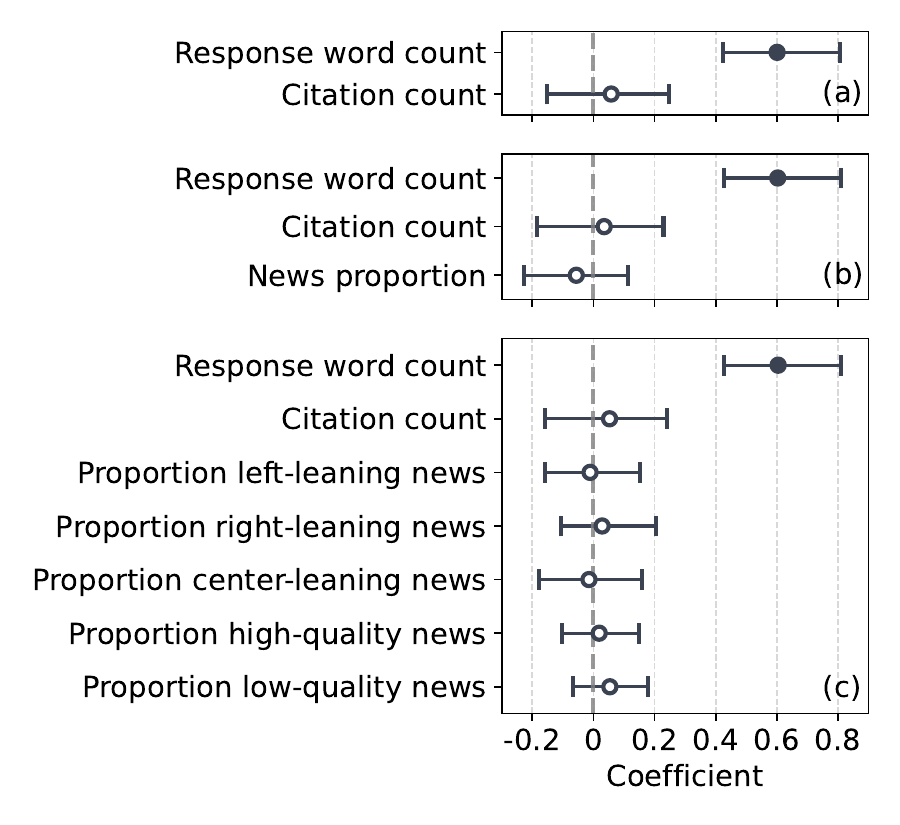}
    \caption{
    Bradley-Terry model coefficients for user preference analysis.
    (a) The base model with only the control variables.
    (b) The model with control variables and proportion of news citations.
    (c) The model with control variables and various news citation features.
    The dots indicate the point estimates and the error bars indicate the 95\% confidence intervals.
    The solid dots indicate statistically significant coefficients at the 0.05 level, while the hollow dots indicate non-significant coefficients.
    }
    \label{fig:user_preference_coefficients}
\end{figure}

To assess the user preference for different news citation features, we show the Bradley-Terry model coefficients in Figure~\ref{fig:user_preference_coefficients}.
We estimate different model specifications by gradually adding news citation features to the base model.

The base model (Figure~\ref{fig:user_preference_coefficients}(a)) includes only control variables: response word count and citation count.
Previous studies have identified these as the most important factors driving user satisfaction~\citep{miroyan2025searcharena}.
When estimated on the complete dataset, both coefficients are positive and significant, confirming prior findings.
However, in our focused analysis of conversations where both AI models cite at least one news source in each response, the significance of the citation count disappears, as shown in the figure.

We then add the proportion of news citations to the base model (Figure~\ref{fig:user_preference_coefficients}(b)).
The coefficient for this variable is not statistically significant, indicating that users do not exhibit a strong preference for the proportion of news citations after controlling for response length and citation count.
Finally, we add various news citation features in terms of political leaning and quality to the base model (Figure~\ref{fig:user_preference_coefficients}(c)).
None of the coefficients are statistically significant, suggesting that users do not have strong preferences for news citations based on political leaning or quality ratings.
The primary factors associated with user preference remain response length.

\section{Discussion}

\subsection{News Citation Patterns and User Preference}

Our analysis provides a comprehensive overview of news citation patterns across AI search models from three major providers using real-world user interactions.
Our findings contribute to the growing literature on AI search systems' potentially disruptive impact on the news ecosystem~\citep{hagar2025ai}.

We reveal rich and varied citation behaviors that address our \textbf{RQ1}.
We find that models within the same family tend to cite similar news sources, even with different configurations.
At the same time, inter-family differences are more pronounced, with different model families relying on substantially different news sources.
Despite these differences, consistent citation patterns emerge across all model families.

We show that news citations are highly concentrated among a small number of well-established outlets---a pattern consistent with previous audits of traditional search engines~\citep{trielli2019search,fischer2020auditing}.
This concentration is most pronounced in OpenAI models, where the top 20 news sources account for 67.3\% of all citations, while Google and Perplexity models show less severe but still substantial concentration.

Another striking finding is the pronounced liberal bias in news citations, with left-leaning and center sources comprising over 98\% of all citations across model families.
This pattern aligns with findings from audits of traditional search engines~\citep{robertson2018auditing,robertson2023users}.
Regarding source quality, all model families demonstrate a clear preference for high-quality outlets, though with notable disparities across providers.
In particular, OpenAI models cite high-quality sources at a rate of 96.2\%, compared to 92.2\% for Google models and 89.7\% for Perplexity models.

Our regression analysis confirms the robustness of these patterns by controlling for various confounding factors.
The coefficients for different question categories reveal that AI search systems adapt their news citation behavior to question types, mirroring patterns observed in traditional search engines~\citep{robertson2018auditing,robertson2023users}.
Surprisingly, however, the results contradict our expectation that AI search systems would cite more politically charged sources when addressing politics-related questions.
Most user questions in our dataset focus on topics that are not inherently political, such as AI and technology or sports and entertainment.
While the ``news updates'' category represents the most likely context for political news citations, its coefficient lacks statistical significance.
These findings suggest that the observed citation patterns stem primarily from the intrinsic characteristics of the AI systems themselves rather than from the nature of user questions.

Our analysis on the user preference addresses our \textbf{RQ2} and confirms findings from previous studies showing that users tend to prefer longer responses in AI search systems~\citep{li2025humantrustaisearch}.
However, neither the proportion of news citations nor the characteristics of cited news sources are associated with user preference with statistical significance.
This finding aligns with the human subject experiment by \citet{li2025humantrustaisearch}, which demonstrates that adding citations increases users' trust in responses regardless of citation validity or relevance.

\subsection{Implications}

Our results confirm that AI search systems already function as powerful algorithmic gatekeepers, with a small number of news outlets capturing the lion's share of citations.
The extreme inequality we observe, particularly for OpenAI models, reveals a winner-take-all dynamic where established brands secure ever-greater visibility while smaller or marginalized outlets struggle to reach audiences through AI-mediated channels, echoing concerns about knowledge sealing in AI search systems~\citep{lindemann2024chatbots}.
Since our Bradley-Terry analysis demonstrates that users prioritize response length over the specific outlets cited, the gatekeeping power of these systems remains largely unchallenged by end-users.

From a democratic perspective, source concentration undermines the diversity of ideas and perspectives available through AI search as gatekeepers~\citep{shoemaker2009gatekeeping}.
This problem is compounded by our finding that concentration occurs alongside significant political bias, meaning users encounter not only fewer sources but sources that lean heavily in one political direction.
The systematic nature of both concentration and political bias is particularly troubling, as these patterns appear consistently across all three AI search providers examined.
This consistency suggests the issue transcends individual system architectures and likely reflects broader patterns in training data, retrieval mechanisms, or optimization objectives.

The variation in concentration patterns and source quality across model families creates a situation where users' access to information depends heavily on their choice of AI system.
For instance, while OpenAI users benefit from citations to higher-quality news sources, they also encounter much greater source concentration compared to users of Google or Perplexity systems.
This disparity may create a new form of digital divide, where the breadth and quality of information access becomes contingent on platform choice or the ability to afford more expensive products.

Our findings on user preference suggest that users may not closely examine the specific sources cited in responses, provided that citations are present.
These results reinforce concerns that AI search systems, by delivering pre-summarized responses, may diminish users' motivation and capacity to explore the broader context surrounding answers and to critically assess the credibility and relevance of cited sources~\citep{shah2022situating,lindemann2024chatbots}.

\subsection{Future Research}

Like many other studies on AI search systems, our study faces some limitations, and the findings should be interpreted accordingly.
Nevertheless, these limitations indicate critical directions for future research.

First, as an observational study of these complex black-box systems, we cannot pinpoint the exact causes of the observed news citation patterns.
AI search systems typically comprise multiple components, including large language models and information retrieval systems~\citep{xiong2024searchengineservicesmeet}.
The mechanisms by which LLMs interpret user queries, retrieve relevant information, and select citation sources could all contribute to the citation patterns.
Therefore, further research is needed to decompose these system components and understand how they individually and collectively influence the final outputs.
We also call for greater industry transparency regarding system architectures and component interactions, as this is essential for enabling public understanding and evaluation of these systems.

Second, our findings underscore the need for a more systematic analysis of AI search systems as they rapidly evolve.
The present study represents a snapshot from a specific time period (March-May 2025) and may not reflect future system performance.
Longitudinal studies tracking changes in citation patterns over time would help determine whether observed biases remain stable or shift with system updates.
Our finding that different platforms exhibit drastically different news citation patterns highlights the importance of including additional providers in future analyses.
It would also be valuable to examine how citation patterns vary across different query types, languages, and contexts beyond our current scope.
Such systematic analysis would not only deepen our understanding of AI search systems' behavior but also enable users to make more informed decisions when selecting among these systems.

Third, our analysis focuses exclusively on news citations, which represent only 9\% of all citations in the dataset.
The remaining 91\% of citations deserve equal scrutiny in future research.
For instance, 10\% of citations point to social media platforms.
Unlike traditional news, this user-generated content presents greater regulatory challenges and creates significant difficulties for both human evaluators and AI systems attempting to assess credibility.
Therefore, social media may serve as particularly problematic vectors for partisan or low-quality information~\citep{shu2017fake}.
Future research should, therefore, examine citation patterns across all source types to develop a comprehensive understanding of how AI search systems curate and prioritize information.

Our findings also raise fundamental questions about the design and governance of AI search systems.
Given the difficulty of quantifying output quality in AI search systems, how can we design evaluation metrics that satisfy both user preferences and the public interest?
This challenge is particularly acute given our finding that these two objectives might not be aligned.
The remedy may require novel approaches to training, retrieval, and ranking that explicitly account for political diversity and source credibility.
Regulatory frameworks may also be needed to ensure that AI search systems serve the public interest by providing balanced and reliable information access.



\bibliography{ref}

\newpage
\appendix

\section{Appendix}

\subsection{Summary Statistics}

\begin{figure}[H]
\centering
\includegraphics[width=\columnwidth]{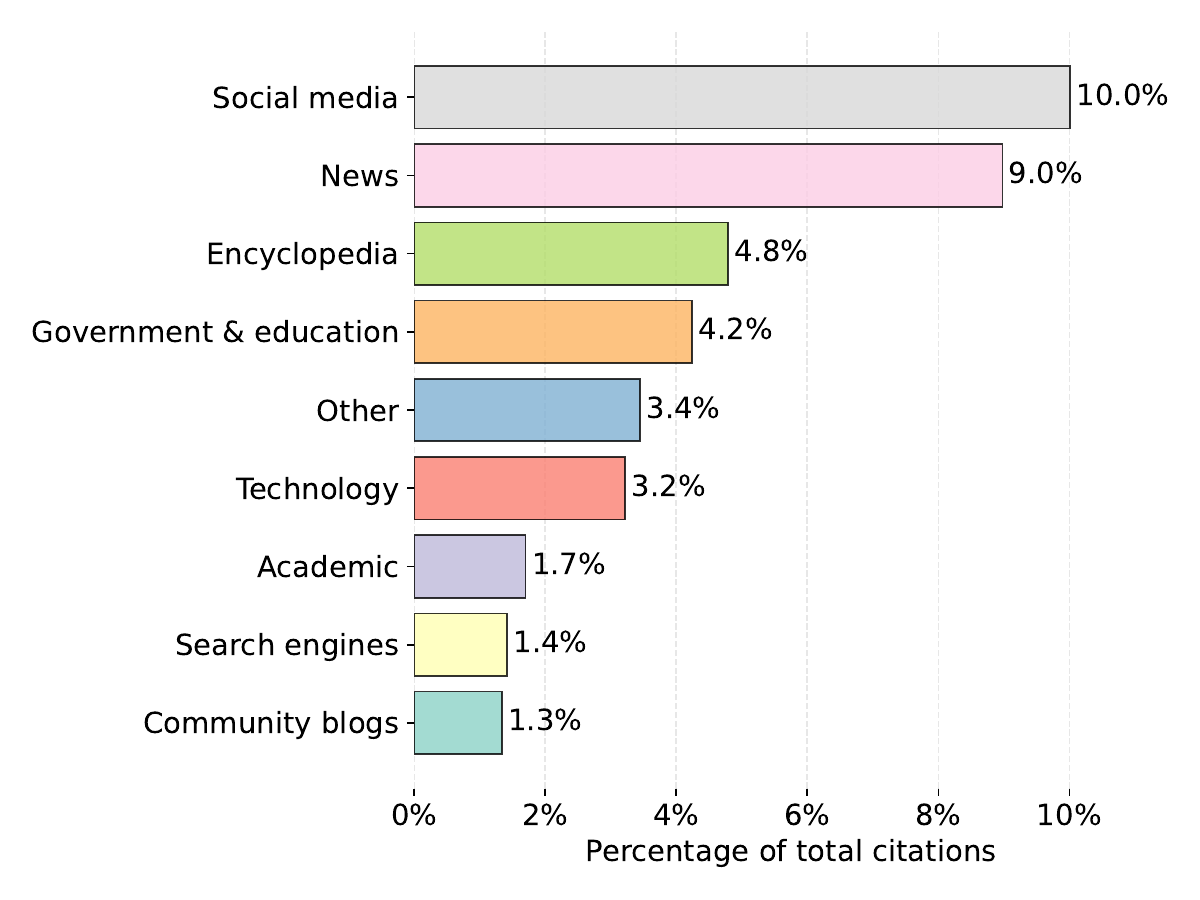}
\caption{
Distribution of citations across domain types.
The ``Unclassified'' category is excluded from the figure but accounts for 60.1\% of all citations.
}
\label{fig:domain_distribution}
\end{figure}

In addition to news domains, we also identify other domain types from the dataset.
We identify government and educational domains based on their domain suffixes and classify them as ``government \& education.''
We also manually identify domains of academic journals (e.g., \domain{science.org}) and research repositories (e.g., \domain{arxiv.org}), classifying them as ``academic.''
Finally, we manually classify the 200 most frequently cited domains into the following categories: ``social media,'' ``encyclopedia,'' ``technology,'' ``search engines,'' ``community blogs,'' and ``other.''
These 200 domains account for 32.6\% of all citations.
All remaining domains are labeled as ``unclassified.''
In Figure~\ref{fig:domain_distribution}, we visualize the distribution of citations across domain types.

\begin{table*}[htbp]
\centering
\caption{
Summary statistics of different models in our dataset.
The table presents the provider, model name, number of responses, total citations, news citations, and the percentage of news citations among all citations in the raw dataset.
The model marked with an asterisk is not a search model and so we exclude it from our analysis.
}
\label{tab:model_stats}
\begin{tabular}{llrrrr}
\toprule
Provider & Model & Responses & Citations & News Citations & News \% \\
\midrule
OpenAI & gpt-4o-mini-search-preview & 3,945 & 7,126 & 1,374 & 19.3\% \\
OpenAI & gpt-4o-search-preview & 5,789 & 9,775 & 1,985 & 20.3\% \\
OpenAI & gpt-4o-search-preview-high & 5,747 & 10,782 & 1,996 & 18.5\% \\
OpenAI & gpt-4o-search-preview-high-loc & 6,995 & 12,354 & 2,288 & 18.5\% \\
\midrule
Perplexity & sonar & 4,513 & 38,266 & 2,794 & 7.3\% \\
Perplexity & sonar-pro & 3,911 & 37,016 & 2,602 & 7.0\% \\
Perplexity & sonar-pro-high & 6,768 & 72,644 & 5,167 & 7.1\% \\
Perplexity & sonar-reasoning-pro-high & 6,455 & 62,235 & 5,662 & 9.1\% \\
Perplexity & sonar-reasoning & 5,941 & 47,497 & 3,601 & 7.6\% \\
\midrule
Google & gemini-2.0-flash-grounding & 5,185 & 17,060 & 1,324 & 7.8\% \\
Google & gemini-2.5-flash-preview-04-17-grounding & 3,359 & 23,886 & 1,583 & 6.6\% \\
Google & gemini-2.5-pro-exp-03-25-grounding & 5,920 & 27,446 & 2,489 & 9.1\% \\
Google & gemini-2.5-pro-exp-03-25-wo-search* & 1,240 & 0 & 0 & 0.0\% \\
\midrule
Total & & 65,768 & 366,087 & 32,865 & 9.0\% \\
\bottomrule
\end{tabular}
\end{table*}

In Table~\ref{tab:model_stats}, we report the summary statistics for the models in our dataset.
Note that the model \model{gemini-2.5-pro-exp-03-25-wo-search} is not a search model.
Although it is included in the raw dataset, we exclude it from our analysis.

Some models in the table may share the same base LLM but differ in configuration parameters.
For instance, \model{gpt-4o-search-preview} and \model{gpt-4o-search-preview-high} both use the \model{gpt-4o} base model, with the latter configured for a higher search context size.
Since models within the same family exhibit similar patterns in our analysis, we aggregate their statistics for most reported results.
For detailed model specifications and configurations, see \citet{miroyan2025searcharena}.

In Table~\ref{tab:model_stats}, we also show the news citation rates across models.
We can see that the news citation rates are similar across models within the same family.
However, OpenAI models tend to cite more news sources in their responses.

\subsection{Model News Citation Similarity}

\begin{figure*}[t]
\centering
\includegraphics[width=0.8\textwidth]{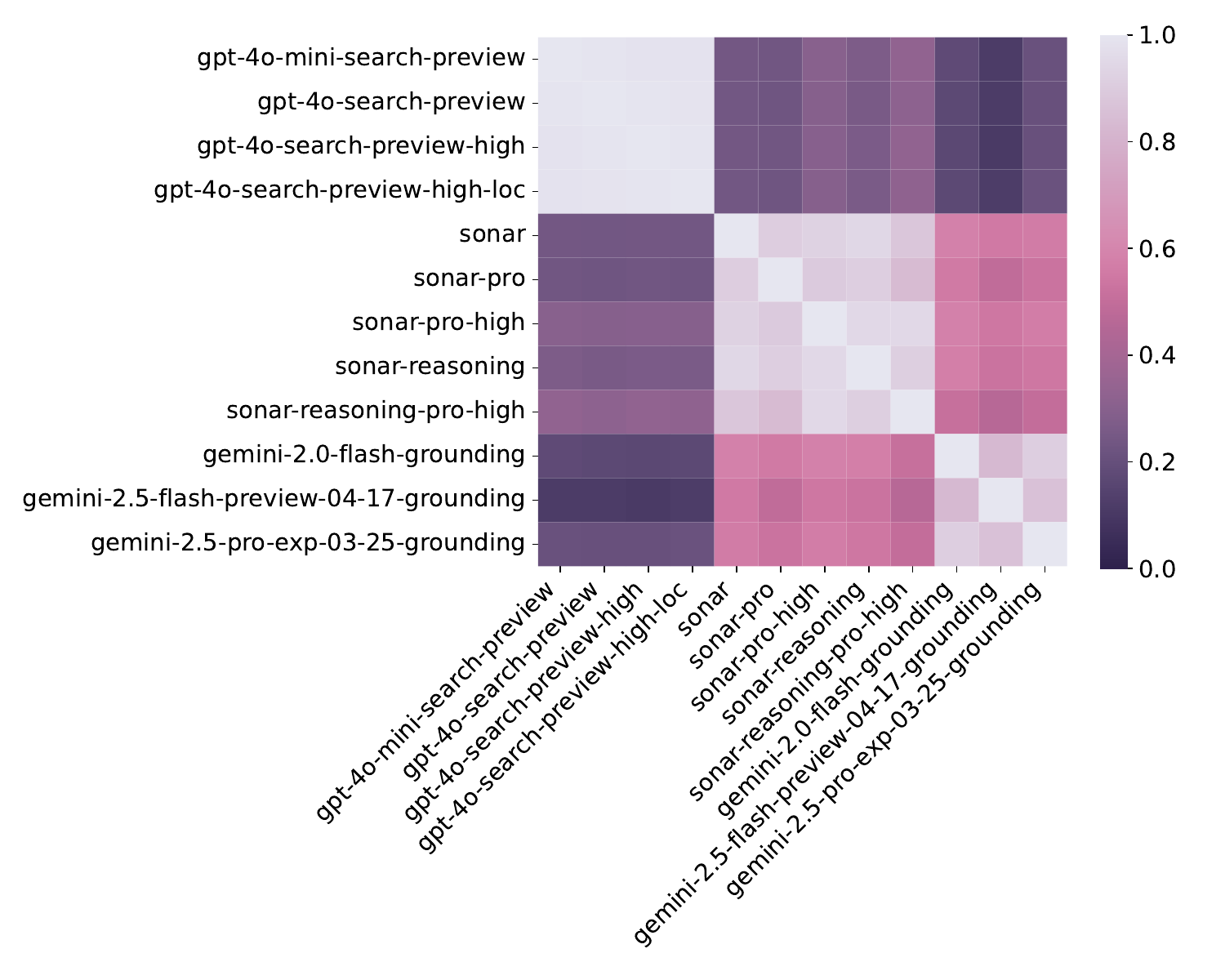}
\caption{Cosine similarity heatmap of cited news domains across models.}
\label{fig:model_news_domain_cosine_similarity_heatmap}
\end{figure*}

Figure~\ref{fig:model_news_domain_cosine_similarity_heatmap} presents a cosine similarity heatmap of cited news domains across models.
The heatmap reveals strong intra-family clustering, with models within each provider family exhibiting highly similar news domain citation patterns.
Examining inter-family differences, we observe that OpenAI models demonstrate distinct citation patterns compared to Google and Perplexity models, with average cosine similarity scores below 0.33.
In contrast, Google and Perplexity models show moderate similarity to each other, achieving average cosine similarity scores of around 0.53.

\subsection{Overrepresented News Domains}

To identify overrepresented domains for a given model family when compared to other model families, we leverage the log-odds ratio informative Dirichlet prior method, originally developed to compare text corpora~\citep{monroe2008fightin}.
This Bayesian framework assumes the frequencies of different news domains follow certain multinomial distributions and incorporates the background information of the typical frequency of each news domain into the model through Dirichlet priors.

The log-odds ratio $\delta_d^{\alpha-\beta}$ of domain $d$ between model family $\alpha$ and model family $\beta$ can be estimated by:

\begin{align}
    \delta_d^{\alpha-\beta} =&& \log{\left( \frac{n^\alpha_d + n^b_d}{N^\alpha + N^b - (n^\alpha_d + n^b_d)} \right)} - \\
    && \log{\left( \frac{n^\beta_d + n^b_d}{N^\beta + N^b - (n^\beta_d + n^b_d)} \right)} \text{,}
\end{align}
where $n^\alpha_d$ ($n^\beta_d$) is the number of citations to domain $d$ in model family $\alpha$ ($\beta$), $n^b_d$ is the number of citations to domain $d$ in the background corpus, $N^\alpha$ ($N^\beta$) is the total number of citations in model family $\alpha$ ($\beta$), and $N^b$ is the total number of citations in the background corpus.

The variance and Z-score of $\delta_d^{\alpha-\beta}$ can be estimated as
\begin{align}
    \sigma^2(\delta_d^{\alpha-\beta}) \approx \frac{1}{n^\alpha_d + n^b_d} + \frac{1}{n^\beta_d + n^b_d} \text{,}
    Z_d^{\alpha-\beta} = \frac{\delta_d^{\alpha-\beta}}{\sqrt{\sigma^2(\delta_d^{\alpha-\beta)}}}.
\end{align}
Positive values of $Z_d^{\alpha-\beta}$ means domain $d$ is over-represented in the cited domains of model family $\alpha$ when compared to model family $\beta$, and the threshold 1.96 can be used to estimate statistical significance with $p < 0.05$.

\begin{table*}[htbp]
\centering
\caption{
    Top 20 overrepresented news sources by model family, ranked by their log-odds ratio Z-scores.
    Political leaning (columns ``L''): L=left-leaning, C=center, R=right-leaning, U=unknown.
    Quality (columns ``Q''): H=high-quality, L=low-quality, U=unknown.
}
\label{tab:overrepresented_news_sources}
\resizebox*{\textwidth}{!}{
\begin{tabular}{lrcc|lrcc|lrcc}
\toprule
\multicolumn{4}{l|}{\textbf{OpenAI}} & \multicolumn{4}{l|}{\textbf{Google}} & \multicolumn{4}{l}{\textbf{Perplexity}} \\
\midrule
Domain & Z-score & L & Q & Domain & Z-score & L & Q & Domain & Z-score & L & Q \\
\midrule
reuters.com & 34.09 & C & H & indiatimes.com & 8.38 & C & L & bbc.com & 9.67 & C & H \\
apnews.com & 22.22 & C & H & livemint.com & 6.62 & C & U & nytimes.com & 8.41 & C & H \\
ft.com & 19.73 & C & H & aljazeera.com & 6.51 & L & H & yahoo.com & 7.70 & C & H \\
axios.com & 18.43 & C & H & thehindu.com & 6.17 & L & H & espn.com & 6.47 & C & H \\
time.com & 11.43 & C & H & dailymail.co.uk & 5.31 & C & L & cnn.com & 6.07 & C & H \\
as.com & 10.63 & C & U & independent.co.uk & 3.79 & C & H & cnbc.com & 5.83 & C & H \\
theatlantic.com & 8.64 & L & H & history.com & 3.58 & C & H & sohu.com & 5.80 & C & L \\
elpais.com & 7.67 & L & H & ndtv.com & 3.58 & C & H & 163.com & 5.77 & C & U \\
theguardian.com & 6.03 & C & H & hindustantimes.com & 3.49 & C & H & economictimes.com & 5.53 & C & U \\
biomedcentral.com & 4.64 & L & H & forbes.com & 3.45 & C & H & techcrunch.com & 5.42 & C & H \\
lemonde.fr & 4.59 & L & H & healthline.com & 3.37 & C & H & nbcnews.com & 5.15 & C & H \\
asahi.com & 2.66 & L & H & cbr.com & 3.20 & C & U & usatoday.com & 4.91 & C & H \\
knowyourmeme.com & 2.58 & L & H & medicalnewstoda... & 2.86 & C & H & rbc.ru & 4.79 & L & U \\
ipsnews.net & 2.58 & L & U & cbsnews.com & 2.78 & C & H & accuweather.com & 4.71 & C & H \\
makeuseof.com & 2.49 & C & H & reliefweb.int & 2.71 & L & U & globo.com & 4.70 & L & U \\
scrippsnews.com & 2.38 & U & U & infobae.com & 2.70 & C & H & npr.org & 4.69 & C & H \\
windowscentral.com & 2.36 & C & U & fool.com & 2.63 & C & H & dw.com & 4.55 & C & H \\
pitchfork.com & 2.11 & L & U & timesofisrael.com & 2.60 & C & H & wsj.com & 4.36 & C & H \\
abc.net.au & 2.11 & C & H & dexerto.com & 2.59 & C & U & euronews.com & 4.10 & C & H \\
bizjournals.com & 2.07 & C & H & techradar.com & 2.59 & C & H & sina.com.cn & 3.82 & L & U \\
\bottomrule
\end{tabular}
}
\end{table*}

For each model family, we compare its citation patterns against the combined patterns of the other two families.
We use the aggregated news domain citation frequencies across all three model families as the background corpus.
Table~\ref{tab:overrepresented_news_sources} presents the top 20 overrepresented news domains for each model family, along with their corresponding log-odds ratio Z-scores.
All reported Z-scores exceed 1.96, indicating statistical significance.

When comparing these overrepresented domains with the most frequently cited sources from the main analysis, we observe that domains commonly cited by multiple model families are filtered out, revealing more distinctive preferences for each model family.
We find that some most frequently cited domains also exhibit high log-odds ratio Z-scores.
For example, OpenAI models disproportionately cite \domain{reuters.com} and \domain{apnews.com}, Google models show a strong preference for \domain{indiatimes.com}, and Perplexity models favor \domain{bbc.com}.

We also include the political leaning and quality categories of the news domains in the table.
We can see that these overrepresented domains are still overwhelmingly left-leaning and high-quality, comfirming the robustness of our main findings.

\subsection{Question Topics}

\begin{table*}
\centering
\caption{
    User question topics discovered through topic modeling by BERTopic.
    Keywords represent the most relevant terms extracted using the KeyBERT algorithm.
    Topic labels provide human-readable descriptions of the discovered topics.
}
\label{tab:question_topics}
\begin{tabular}{cp{10cm}l}
\toprule
Topic & Keywords & Label \\
\midrule
0 & ai, ai model, models, model, google, API, search, web, data, code & AI models and technology \\
1 & stock price, stock market, stock, stocks, market, investor, markets, invest, trading, investment & stock prices and market \\
2 & healthy, fat, diabetes, supplements, vegan, foods, health, eat, diseases, disease & diet, nutrients, and health \\
3 & news latest, latest news, news today, today news, news, world news, latest, tell latest, recent, today & news updates \\
4 & matches, final, tournament, South America, teams, won, match, winners, champions, continent & sports and entertainment \\
5 & software engineer, engineer, graduated, career, actor, Bucharest, contact, mastodon, software, university & biography and personal stories \\
6 & battle, battles, combat, strongest, vs, abilities, weapon, characters, weapons, main & fictional character battle analysis \\
7 & book, chapter, chapters, summary, summary latest, authors, wrote, read, old testament, testament & online content and book \\
8 & lyrics, song, testimony, plead, songs, son, minor, rap, artist, tune & music and lyrics \\
9 & tails, sonic, villain, spider, murderer, killed, evil, chaos, tyrant, nickname & comics and games \\
\bottomrule
\end{tabular}
\end{table*}

In Table~\ref{tab:question_topics}, we present the full list of question topics discovered through topic modeling and their corresponding keywords and labels.
The keywords are identified using KeyBERT algorithm~\citep{grootendorst2020keybert}.

\subsection{Regression Analysis Results}

\begin{table*}
\centering
\caption{
    Regression coefficients for news source citation patterns.
    Coefficients show the relationship between features and news source citation patterns.
    Statistical significance: *** $p < 0.001$, ** $p < 0.01$, * $p < 0.05$.
    Sample size: 9,098 observations.
}
\label{tab:regression_coefficients}
\resizebox*{\textwidth}{!}{
\begin{tabular}{lrrrr}
    \toprule
Feature & \% left-leaning news & \% right-leaning news & \% high quality news & \% low quality news \\
\midrule
Intercept & 18.53*** & 2.46*** & 69.93*** & 5.69*** \\
Model family: Google & 3.95*** & -0.92** & -4.97*** & 2.26*** \\
Model family: Openai & -2.76** & -0.37 & 2.31* & -0.73 \\
Number of citations & -0.19* & -0.04 & 0.04 & -0.08 \\
News sources percentage & -0.06*** & -0.02*** & 0.16*** & -0.03*** \\
Response length (words, log Z-score) & -0.23 & 0.75*** & 1.41 & 0.13 \\
Total turns & -0.07 & -0.02 & -0.18 & 0.10 \\
Turn number & 0.98** & -0.05 & 0.47 & -0.11 \\
Question length (words, log Z-score) & -0.42 & 0.23 & -0.10 & 0.32 \\
Client country/region: BR & -0.08 & 2.30* & -6.89* & -1.03 \\
Client country/region: CA & 3.36* & -1.07 & 3.71* & -1.02 \\
Client country/region: CN & 7.53* & 1.61 & -0.29 & 3.29* \\
Client country/region: DE & 3.05 & -0.28 & 4.25* & -0.62 \\
Client country/region: GB & 0.12 & -0.04 & 1.96 & -0.18 \\
Client country/region: HK & -4.25 & -1.02 & 7.08* & -0.25 \\
Client country/region: IN & 1.40 & -0.56 & -6.65*** & 9.78*** \\
Client country/region: MX & 10.29*** & 5.54*** & 5.58* & 1.05 \\
Client country/region: other & 1.73 & -0.52 & -0.11 & -0.29 \\
Client country/region: RU & 10.21*** & -0.89 & -0.46 & 0.27 \\
Client country/region: unknown & 6.15*** & 0.32 & -2.37* & -0.43 \\
Embedding: PC 0 & -5.69*** & -0.70*** & 2.25*** & -0.90*** \\
Embedding: PC 1 & 0.80 & 0.10 & 4.91*** & 0.83*** \\
Embedding: PC 10 & 1.07** & -0.41*** & -3.60*** & 0.40* \\
Embedding: PC 11 & -0.08 & -0.11 & 1.81*** & -0.13 \\
Embedding: PC 12 & 0.08 & 0.40** & 0.82* & 0.14 \\
Embedding: PC 13 & -0.61 & 0.30* & -0.89* & 0.50** \\
Embedding: PC 14 & 2.60*** & 0.12 & 0.55 & -0.04 \\
Embedding: PC 15 & 1.00** & -0.08 & 0.41 & -0.63*** \\
Embedding: PC 16 & 0.26 & 0.34** & 0.36 & -0.08 \\
Embedding: PC 17 & 0.07 & 0.58*** & -2.09*** & -0.82*** \\
Embedding: PC 18 & 0.02 & 0.20 & 0.24 & -0.35* \\
Embedding: PC 19 & 0.73* & -1.40*** & 1.71*** & -0.61*** \\
Embedding: PC 2 & -1.89*** & -0.02 & -2.37*** & -0.07 \\
Embedding: PC 3 & -5.16*** & -0.54*** & -2.54*** & 0.10 \\
Embedding: PC 4 & -0.74 & -0.09 & -3.45*** & -0.81*** \\
Embedding: PC 5 & 1.05** & -0.25* & 0.93* & -0.02 \\
Embedding: PC 6 & 0.80* & 0.32** & -4.05*** & 0.01 \\
Embedding: PC 7 & -0.76* & -0.31* & 0.67 & -0.72*** \\
Embedding: PC 8 & -0.47 & 0.21 & -1.11** & -0.54** \\
Embedding: PC 9 & -1.68*** & -0.40** & 0.50 & 0.24 \\
Intent: analysis & -1.99 & 0.58 & -1.08 & -0.39 \\
Intent: creative generation & -0.02 & -2.26*** & -1.75 & -1.80 \\
Intent: explanation & -0.27 & -0.35 & -1.39 & 0.16 \\
Intent: factual lookup & 0.69 & 0.20 & -2.85 & -0.61 \\
Intent: guidance & -6.49*** & -1.07 & -6.42** & -1.30 \\
Intent: info synthesis & -1.70 & 0.89 & -2.37 & -1.41 \\
Intent: other & -1.24 & 0.22 & 2.84 & -1.17 \\
Intent: text processing & -0.77 & 3.68** & -0.56 & -2.13 \\
Topic: AI models and technology & -0.58 & 0.02 & 2.42*** & -0.34 \\
Topic: stock prices and market & 0.37 & 0.11 & -1.21*** & 0.15 \\
Topic: diet, nutrients, and health & 0.11 & 0.04 & 0.90* & -0.09 \\
Topic: news updates & -0.43 & -0.14 & -0.06 & -0.16 \\
Topic: sports and entertainment & -0.30 & 0.17 & -0.05 & 0.26 \\
Topic: biography and personal stories & 0.14 & 0.22* & 0.01 & 0.24 \\
Topic: fictional character battle analysis & 0.23 & -0.12 & -2.17*** & 0.00 \\
Topic: online content and book & 0.38 & -0.78*** & 1.64*** & -0.42 \\
Topic: music and lyrics & 0.05 & 0.27 & -1.82*** & 0.02 \\
Topic: comics and games & 0.04 & -0.20 & -0.47 & 0.20 \\
\midrule
$R^2$ & 0.12 & 0.06 & 0.13 & 0.05 \\
\bottomrule
\end{tabular}
}
\end{table*}

In Table~\ref{tab:regression_coefficients}, we present the full linear regression analysis results.

\end{document}